\pdfoutput=1
\documentclass[cernpreprint,english]{na61doc}

\usepackage{subcaption}
\usepackage{multicol}
\usepackage[shortlabels]{enumitem}

\usepackage{colortbl}
\definecolor{darkred}{rgb}{0.5,0,0}
\definecolor{darkblue}{rgb}{0,0,0.5}
\definecolor{firebrick}{rgb}{0.75,0.125,0.125}
\definecolor{darkgreen}{rgb}{0,0.5,0}

\usepackage{cite}
\usepackage[colorlinks=true,linkcolor=firebrick,citecolor=darkgreen,urlcolor=darkblue]{hyperref}

\newcommand{\eV}{\ensuremath{\mbox{e\kern-0.1em V}}\xspace}
\newcommand{\GeV}{\ensuremath{\mbox{Ge\kern-0.1em V}}\xspace}
\newcommand{\MeV}{\ensuremath{\mbox{Me\kern-0.1em V}}\xspace}
\newcommand{\GeVc}{\ensuremath{\mbox{Ge\kern-0.1em V}\!/\!c}\xspace}
\newcommand{\GeVcc}{\ensuremath{\mbox{Ge\kern-0.1em V}\!/\!c^2}\xspace}
\newcommand{\AGeV}{\ensuremath{A\,\mbox{Ge\kern-0.1em V}}\xspace}
\newcommand{\AGeVc}{\ensuremath{A\,\mbox{Ge\kern-0.1em V}\!/\!c}\xspace}
\newcommand{\MeVc}{\ensuremath{\mbox{Me\kern-0.1em V}/c}\xspace}

\newcommand{\dd}{\ensuremath{{\mathrm d}}\xspace}
\newcommand{\dedx}{\ensuremath{\dd E\!/\!\dd x}\xspace}

\newcommand{\FlukaEleven}{{\scshape Fluka2011}\xspace}
\newcommand{\Fluka}{{\scshape Fluka}\xspace}

\newcommand{\GeantFour}{{\scshape Geant4}\xspace}

\newcommand{\CernVM}{\textsc{Cern\-\kern-0.05emVM}\xspace}

\ShineTitle{Measurement of the production cross section of 31~\GeVc protons on carbon via beam attenuation in a 90-cm-long target}

\PreprintIdNumber{CERN-EP-2020-198}

\ShineJournal{Phys. Rev. D}
\ShineAbstract{
The production cross section of 30.92~\GeVc protons on carbon is measured by the \NASixtyOne spectrometer at the CERN SPS by means of beam attenuation in a copy~(replica) of the 90-cm-long target of the T2K neutrino oscillation experiment. The employed method for direct production cross-section estimation minimizes model corrections for elastic and quasi-elastic interactions. The obtained production cross section is $\sigma_\mathrm{prod}~=~227.6~\pm~0.8\mathrm{(stat)}~_{-~3.2}^{+~1.9}\mathrm{(sys)}~{-~0.8}\mathrm{(mod)}$~mb. It is in agreement with previous \NASixtyOne results obtained with a thin carbon target, while providing improved precision with a total fractional uncertainty of less than 2$\%$. This direct measurement will reduce the uncertainty of the T2K neutrino flux prediction.
}
\begin{document}
\maketitle

\section{Introduction}
\begin{figure}
    \centering
    \includegraphics[width=0.5\textwidth, height=2cm]{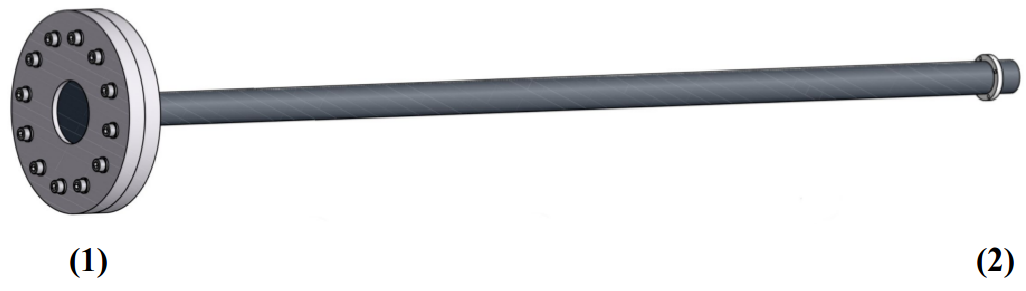}
\caption{The T2K replica target with the upstream aluminum flange (1) and the downstream plastic ring (2) used to support the target.}
    \label{fig:target}
\end{figure}

\NASixtyOne (SPS Heavy Ion and Neutrino Experiment)
\cite{Abgrall:2014fa} is a fixed-target experiment at the CERN Super Proton Synchrotron (SPS). It has a rich physics program covering three different fields of research: strong-interaction physics, cosmic ray physics and neutrino physics. \NASixtyOne's neutrino program includes hadron production measurements for the Tokai-to-Kamioka (T2K) long-baseline neutrino oscillation experiment in Japan~\cite{Abe:2011ks} and Fermilab neutrino experiments on the NuMI and LBNF beam lines:  MINER$\nu$A~\cite{Aliaga:2013uqz}, NO$\nu$A~\cite{Ayres:2007tu} and DUNE~\cite{Acciarri:2015uup}. These measurements are used to reduce the systematic uncertainties associated with the prediction of the corresponding (anti)neutrino fluxes. A new calculation of the production cross section for protons on carbon \mbox{(p + C)} via a measurement of the number of beam particles remaining after interactions in a replica of the 90-cm-long T2K graphite target is presented here.\\
The \NASixtyOne collaboration uses the following classification of nuclear interactions based on the type of particles produced in the process\footnote{The same definitions of nuclear interaction types are used by Bellettini \textit{et al.} in ~\cite{Bellettini:1966zz}, where measurements of proton-nuclei cross-sections are described and interpreted.}. An elastic process, that is coherent nuclear elastic scattering, is one in which no new particles are produced. Except for coherent nuclear scattering, all other processes due to the strong interaction are labeled as inelastic. Furthermore, the inelastic interactions are subdivided into two groups: quasi-elastic and production processes. Quasi-elastic processes result in the fragmentation of the target nucleus. In production interactions, new hadrons are produced. The probability for each of these three processes is governed by the corresponding cross section and the total cross section is:
\begin{eqnarray}
    \sigma_\mathrm{tot} =\sigma_\mathrm{el}+\sigma_\mathrm{inel},
\end{eqnarray}
where
\begin{eqnarray}
    \sigma_\mathrm{inel} =\sigma_\mathrm{qe}+\sigma_\mathrm{prod}.
\end{eqnarray}
The \NASixtyOne experiment collected hadron production data for T2K during the runs in years 2007, 2009 and 2010. A 30.92~\GeVc proton beam and both thin and thick (90-cm-long T2K replica) targets were used. Primary hadron interactions in the target were examined using a 2-cm-thin carbon target, with a thickness equal to 4$\%$ of a nuclear interaction length, $\lambda_{I}$. Both single and multiple re-interactions inside the target were simultaneously studied using a long graphite target which is equivalent to about 2$\lambda_{I}$. A schematic view of this target is shown in Fig.~\ref{fig:target}. It is a copy of the T2K target - a 90-cm-long cylinder with a 1.3-cm radius. Previously obtained results from the \NASixtyOne hadron production measurements can be found in~\cite{Abgrall:2012pp,Abgrall:2011ae,Abgrall:2011ts,Abgrall:2013wda,Abgrall:2015hmv,Abgrall:2016jif,Abgrall:2019p}. They include inelastic and production cross-section measurements from the thin target runs and hadron yields from both the thin and replica target data sets. Their full incorporation in the T2K neutrino flux simulation is ongoing. However, it has already been reported that thin target results reduced the systematic uncertainty of the (anti)neutrino fluxes in T2K to about 10$\%$~\cite{Abe:2012av}, and further reduction to around 5$\%$ was achieved after incorporating the 2009 replica target charged pion results~\cite{Zambelli:2017fl}. Even stronger constraints (about 4$\%$) are expected after including the measured charged-pion, kaon and proton yields from the 2010 replica target data. 

The focus of this paper is a measurement with the T2K replica target from the 2010 run, namely the extraction of the production cross section based on the number of beam protons that scattered elastically or quasi-elastically in the long target. This result is to be included in the future in the hadron interaction rate re-weighting procedure that is part of the T2K neutrino flux prediction. The goal of the current measurement is to improve the precision of the previously obtained production cross-section results from the 2007 and 2009 \NASixtyOne thin target data~\cite{Abgrall:2011ae,Abgrall:2015hmv}, thus reducing the associated systematic uncertainty on the T2K flux prediction.

The structure of this paper is as follows: An overview of the \NASixtyOne spectrometer is given in Sec.~\ref{sec:na61}. Section~\ref{sec:ana} describes the analysis procedure. The estimation of systematic uncertainties is presented in Sec.~\ref{sec:sys}, followed by a discussion of the Monte Carlo models in Sec.~\ref{sec:model}. Section~\ref{sec:res} reports the resulting production cross section and compares it to previous measurements. A brief conclusion closes the paper. 
\section{The \NASixtyOne experiment}
\label{sec:na61}
\label{sec:na61det}
\begin{figure}[t!]
    \begin{subfigure}[t]{0.99\textwidth}
        \centering
        \includegraphics[width=\linewidth]{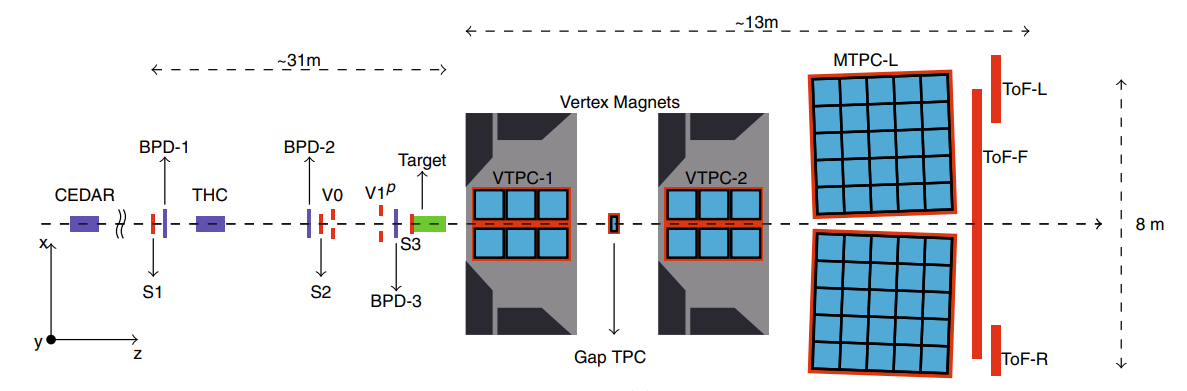}\\
        \caption{}
        \label{fig:na61setup_top}
    \end{subfigure}
    \begin{subfigure}[t]{0.99\textwidth}
        \centering
        \includegraphics[width=\linewidth]{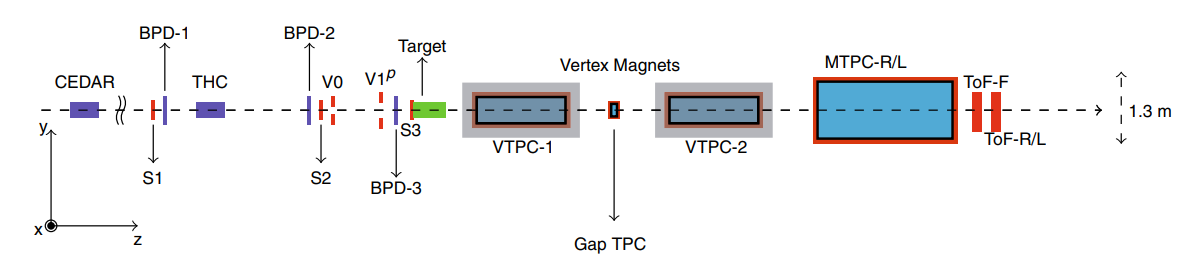}\\
        \caption{}
        \label{fig:na61setup_side}
    \end{subfigure}
\caption{Top (\subref{fig:na61setup_top}) and side (\subref{fig:na61setup_side}) view of the \NASixtyOne experimental setup  during the 2010 T2K replica target data-taking. The orientation of the coordinate system is shown in the bottom-left corner. The beam is aligned with the $z$-axis and comes from the left.}
\label{fig:na61setup}
\end{figure}

The \NASixtyOne experiment is located on a secondary beam line, the H2 beam line, of the CERN SPS. The 400~\GeVc primary SPS proton beam first strikes a target 535 m upstream of the experiment to produce a secondary beam. Then, along the H2 line, a magnet system is employed to select the desired particle momentum according to particle rigidity. Further beam identification is performed by the detectors of the \NASixtyOne trigger system. \\
A schematic view of the \NASixtyOne setup and its coordinate system is given in Fig.~\ref{fig:na61setup}. The beam direction and size are checked by a set of scintillator counters - \mbox{S1, S2} and \mbox{S3} used in coincidence and \mbox{$V_{0}$} and \mbox{$V_{1}^{P}$} used in anti-coincidence. The \mbox{S1} counter provides a timing reference (start signal) for the experiment. Sitting just 0.5 cm upstream of the target, the S3 counter selects particles that hit the target. Both veto counters, \mbox{$V_{0}$} and \mbox{$V_{1}^{P}$}, have 1-cm holes centered around the beam line and allow for selection of a narrow beam. Beam particle identification is done by two Cherenkov detectors - Cherenkov Differential Counter with Achromatic Ring Focus~\cite{CEDAR} (\mbox{CEDAR}) and Threshold Cherenkov Counter (\mbox{THC}). For particles of the same momentum,  only a certain particle type will produce Cherenkov light, depending on the gas pressure in the Cherenkov detectors. This allows for particle-type selection and can provide an estimate of the beam composition at a given energy. 
The trajectory of every incoming beam particle is reconstructed by a set of three Beam Position Detectors:  \mbox{BPD-1, 2} and 3. Each BPD consists of two orthogonal multi-wire proportional chambers that are comprised of a sense wire plane and a cathode strip plane. The BPDs measure the position of the beam particle in the transverse plane with a precision of~200~$\mu$m~\cite{Abgrall:2014fa}.\\
After the beam hits the target, the resulting particles go through the \NASixtyOne spectrometer. During the 2010 run it comprised five Time Projection Chambers (TPC) and three Time-of-Flight (\mbox{ToF}) scintillator walls. Two of the TPCs, Vertex TPC1 (\mbox{VTPC-1}) and Vertex TPC2 (\mbox{VTPC-2}), are placed inside two superconducting dipole magnets. Their combined maximum bending power is 9~T$\thinspace$m. Both~\mbox{VTPC-1} and \mbox{VTPC-2} have one section on the left and one on the right side of the beam line forming a 24-cm-wide gap centered around the beam. A third TPC, the Gap TPC (\mbox{GTPC}), covers the forward region standing between the two Vertex TPCs and is in the residual magnetic field of the two dipole magnets. Further downstream of \mbox{VTPC-2} are Main TPC Left (\mbox{MTPC-L}) and Main TPC Right (\mbox{MTPC-R}). \NASixtyOne's TPC system provides precise momentum, charge and specific energy loss (\dedx) measurements. Additional particle identification, below 8~\GeVc, is done with the Time-of-Flight \mbox{detectors}:~two side ToF-Left~(\mbox{ToF-L}) and ToF-Right (\mbox{ToF-R}) walls and a third ToF-Forward (\mbox{ToF-F}) wall. The \mbox{ToF-F} detector is crucial for the T2K hadron production measurements. It consists of 80 plastic scintillator bars oriented vertically and arranged in ten separate modules. The dimensions of each bar are \mbox{Width $\times$ Height $\times$ Length = 10 $\times$ 120 $\times$ 2.5~cm$^{3}$}. Two photomultiplier tubes (PMTs) on the top and the bottom of each bar read out the produced signal. 
\section{Analysis}
\label{sec:ana}
\subsection{Data collected}
During the year 2010 data-taking period, \NASixtyOne carried out two types of replica target runs. The target was placed at the entrance of the TPC system, longitudinally aligned along the $z$-axis. The majority of events, 9~$\times 10^{6}$, were collected using a 1.2~T$\thinspace$m magnetic field configuration - also referred to as the "low" magnetic field. These events were used to measure the hadron yields from the surface of the T2K replica target, and the corresponding results were published in~\cite{Abgrall:2019p}. In between the low-magnetic field runs, the dipole magnets were operated at their maximum 9~T$\thinspace$m bending power and another 1.2~$\times 10^{6}$~events were recorded. This allowed for better detection of beam protons that scattered elastically or quasi-elastically and passed through the target. Analysis of this maximum magnetic field data set is the subject of this paper.
\subsection{Target position and alignment}
\label{subsec:Tpos}
A dedicated study of the target position,  target tilt, its alignment with respect to the BPDs,  and the BPD-TPC alignment was carried out for the analysis of the low magnetic field replica target data and it is described in detail in~\cite{Abgrall:2019p}. The resulting values and their uncertainties are given in Tab.~\ref{tab:position} and were adopted here for the maximum magnetic field data analysis.\\
\begin{table}[ht]
\centering
\begin{tabular}{ccccccc}
\hline \hline
     &   & x (cm)& y (cm) &z (cm) & $t_{x}$ (mrad) & $t_{y}$ (mrad)\\
    \hline
     & Value & 0.15 &0.12&-657.5 &0.0 & 0.0 \\
     & Uncertainty & 0.03& 0.02 & 0.1& 0.3&0.3\\
\hline \hline
\end{tabular}
\caption{Target upstream position and target tilt in the NA61/SHINE coordinate system (see Fig.~\ref{fig:na61setup}) and their uncertainties~\cite{Abgrall:2019p}. The target tilt is a measure of the longitudinal alignment of the target to the nominal beam direction, the $z$-axis, and is given separately for the $xz$-plane ($t_{x}$) and the $yz$-plane ($t_{y}$).}
\label{tab:position}
\end{table}
\subsection{Event selection}
\label{subsec:event_sel}
\begin{figure}
    \centering
    \includegraphics[width=0.46\textwidth, height=0.44\textwidth]{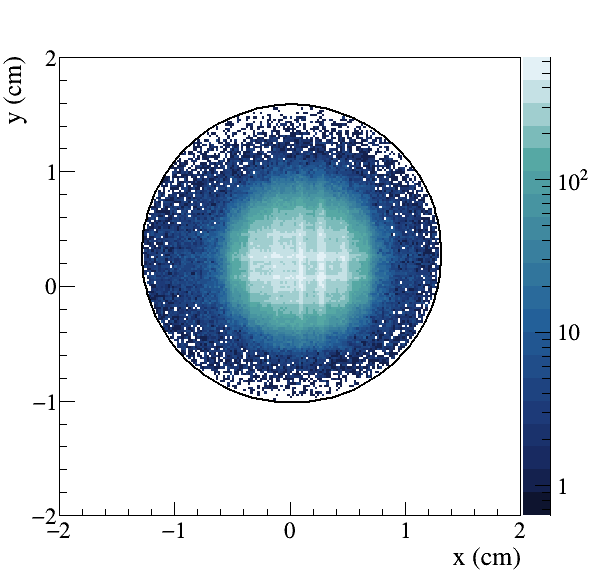}
\caption{Beam profile at the upstream target face with respect to the BPD measurements. The black circle represents the target upstream contour and has a 1.3-cm radius. The mesh structure originates from the finite width and pitch between the strips on the cathode planes of the BPDs.}
    \label{fig:beam_profile}
\end{figure}

To maximize the precision of the beam track reconstruction, each selected beam particle is  required to be detected in all three BPDs. To ensure that the proton beam strikes the target and to reduce systematic biases caused by beam divergence, all events must also satisfy the so-called "T3 trigger" condition.  This corresponds to the following signals in the scintillator counters and the Cherenkov detectors as described in Sec.~\ref{sec:na61det}:
\begin{center}
    T3 = S1 $\: \wedge \:$ S2 $ \:\wedge\: \overline{V_{0}} \:\wedge\: \overline{V}^{p}_{1} \:\wedge\:$ \textrm{CEDAR} $\:\wedge\: \overline{\mathrm{THC}}$\\
\end{center}
Due to the strong magnetic field and because the 90-cm-long replica is about 67 cm upstream of \mbox{VTPC-1} and the upstream magnet, the fringe magnetic field in the target area is non-negligible. Therefore, operation of the scintillator counter closest to the target, the \mbox{S3} counter, was affected by the magnetic field and the signal from this scintillator was removed from the trigger. \\
Event selection is finalized by a requirement that the reconstructed path of the beam particle must have passed through the whole target length. The mean proton beam divergence during the data taking was about 0.3~mrad. This final cut removes events where the beam particle hits the target close to its edge and at a relatively large angle. With full event selection applied, the ($x-y$) profile of selected beam tracks extrapolated to the $z$ position of the target upstream face is shown in Fig.~\ref{fig:beam_profile}, where the black circle indicates the boundary of the target front-face area. Since in T2K the beam profile changes on a run-by-run basis, perfect agreement between the \NASixtyOne beam properties for a given hadron production measurement and the T2K beam is not feasible. However, a comparison of the radial distributions of the two beams in given runs can be found in Fig.~3 in Ref.~\cite{Abgrall:2019p}, where the T2K beam profile is wider than the used T3 beam profile.
\subsection{Track selection}
\label{subsec:track_sel}
\begin{figure}[t!]
    \begin{subfigure}[t]{0.99\textwidth}
        \centering
        \includegraphics[width=0.6\linewidth, height=4.75cm]{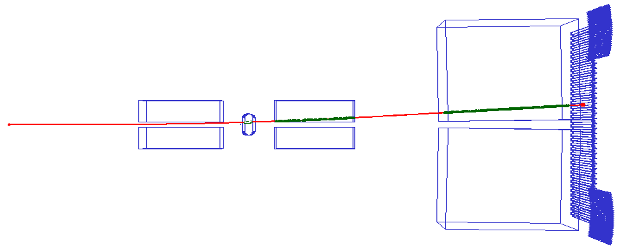}\\
        \caption{}
        \label{fig:pile_up_xz}
    \end{subfigure}
    \begin{subfigure}[t]{0.99\textwidth}
        \centering
        \includegraphics[width=0.6\linewidth, height=4.6cm]{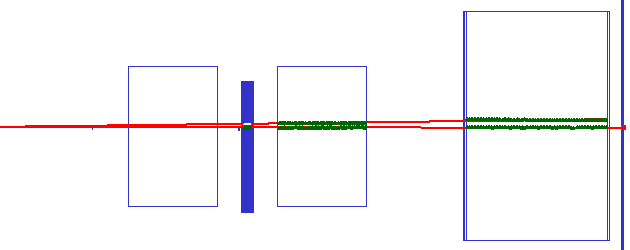}\\
        \caption{}
        \label{fig:pile_up_yz}
    \end{subfigure}
\caption{Top~(\subref{fig:pile_up_xz}) and side~(\subref{fig:pile_up_yz}) view of the tracks of two high-energy particles going through the TPCs. The beam comes from the left. The track points are shown in green and the reconstructed TPC tracks are given in red. The two tracks are indistinguishable in the $xz$-plane shown on plot~\subref{fig:pile_up_xz}. Plot~\subref{fig:pile_up_yz}, illustrating the $yz$-plane, shows that only one of the particles is detected by the \mbox{ToF-F} wall on the right. This is marked by a big red dot. The other track is of an off-time particle that scattered elastically or quasi-elastically in the target and did not produce a signal in the \mbox{ToF} detector (busy processing the previous hit). The $y$ position of the \mbox{ToF-F} hit is reconstructed based on the timing difference between the signals of the two PMTs at the ends of the hit scintillator bar.}
\label{fig:off_time}
\end{figure}

\begin{figure}
    \centering
    \includegraphics[scale=0.35]{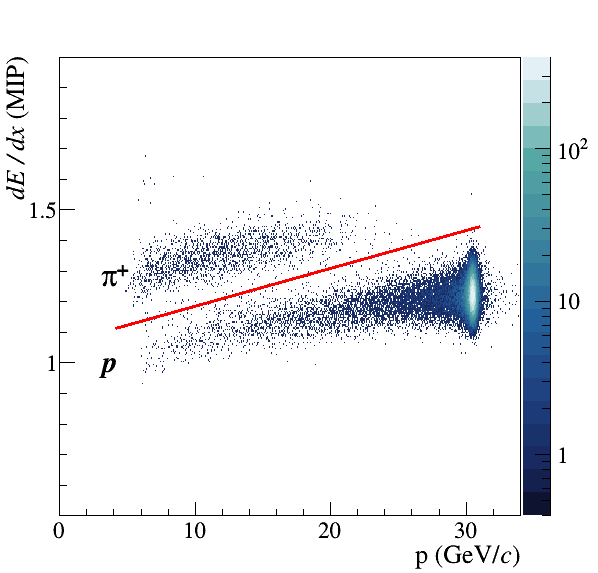}
\caption{Distribution of energy loss in the TPCs as a function of momentum. Good separation between $\pi^{+}$ and protons is observed. The red line shows the graphical cut applied in the track selection.}
    \label{fig:dedx}
\end{figure}

\begin{figure}
    \centering
    \includegraphics[scale=0.21]{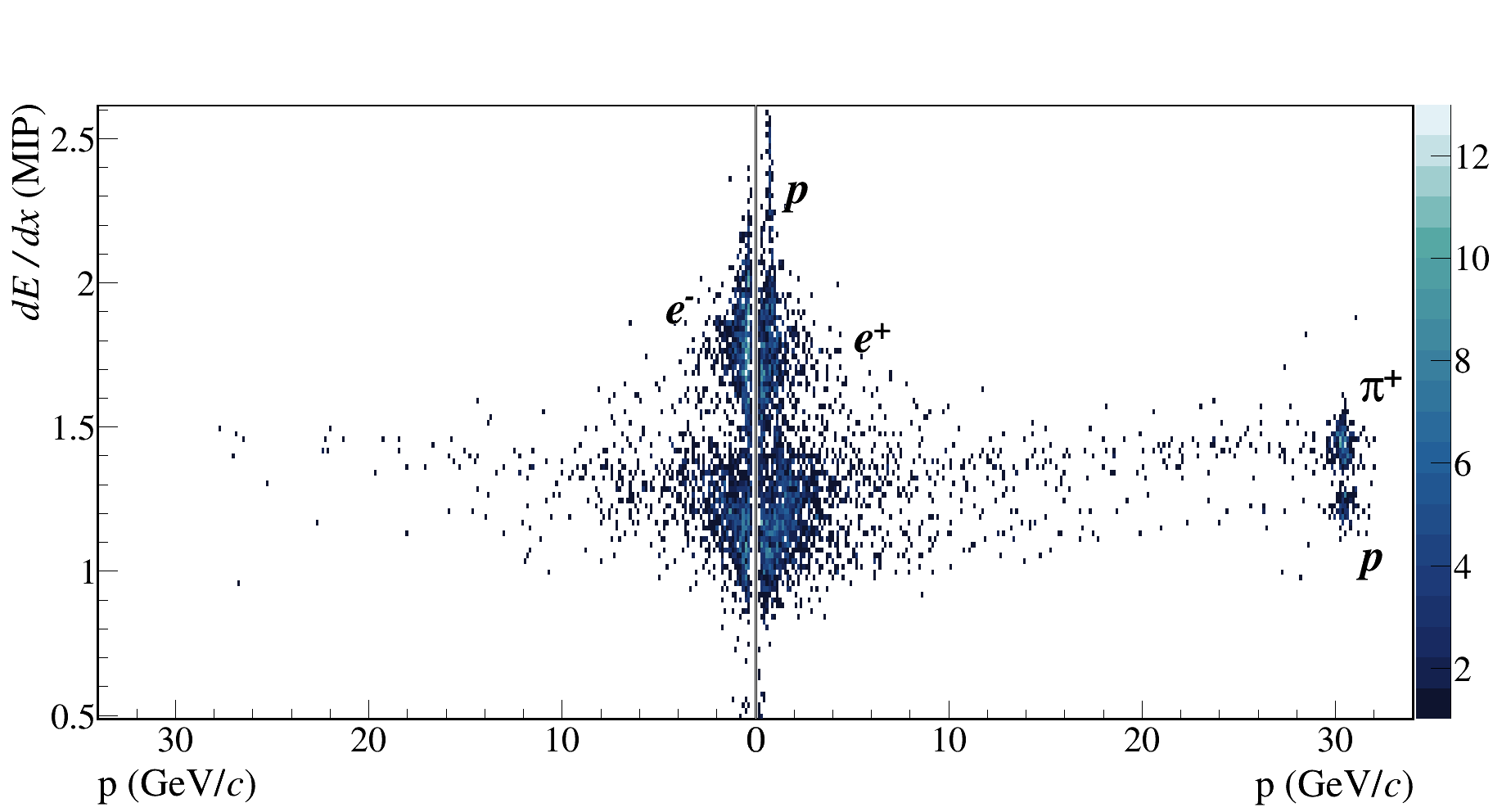}
\caption{Distribution of energy loss in the TPCs as a function of momentum for tracks produced alongside the high-energy selection candidate. On the left is the distribution for negatively charged particles, q<0, while on the right is the distribution for positively charged ones, q>0. In the low-momentum regions on both plots separation between particle species is troublesome. For positive particles at momenta around the 30.92~\GeVc proton beam momentum, there are two distinct peaks. These are off-time beam particles that have survived the nuclear interactions inside the target -  off-time protons (lower) and $\pi^{+}$ (higher) in \dedx.}
    \label{fig:dedx_others}
\end{figure}

For all of the particles that go through the \NASixtyOne TPCs, track selection is designed to extract high-energy beam particles that have passed through the whole target without producing any new hadron.\\
Track reconstruction in the \NASixtyOne software framework requires that each track must have more than 5 clusters in at least one TPC. Furthermore, tracks are selected only if they have a reconstructed momentum. Then, they are extrapolated backwards from the TPCs to the target surface. Extrapolation continues until the track hits the downstream target face or the minimum distance between the target body surface is reached. If this distance is less than three times the radial uncertainty on the extrapolated position, the track is selected. This requirement removes most particles that do not come from the target, but are products of decays or interactions outside the target. Also, only positively charged particles that exit the target from the downstream target face are analyzed.\\
The removal of off-time TPC tracks is ensured by a requirement that tracks must be detected by the \mbox{ToF-F} wall. Such functionality of the \mbox{ToF-F} detector is possible since its acquisition time window is 100~ns, while during the data-taking period the mean time difference between beam particles was around 120~$\mu$s and the readout time of the TPCs is 50~$\mu$s. An example of a pile-up event is given in Fig.~\ref{fig:off_time}, where tracks of two particles are reconstructed by the TPCs in the same readout period. Both particles passed through the target without producing new hadrons. Their reconstructed momentum is approximately the same and is also close to the proton beam momentum of 30.92~\GeVc. Momentum conservation requires that one of these two particles is a pile-up event. The particle that first entered the spectrometer passed the event selection and had a recorded \mbox{ToF-F} hit. The second beam particle, the off-time particle, came afterwards and was not detected by the \mbox{ToF-F} wall as it hit the same \mbox{ToF-F} scintillator bar while the signal of the first particle was being processed. Hence, the \mbox{ToF-F} hit requirement allows for the rejection off-time TPC tracks. The case of an off-time beam particle that produces new hadrons in the target is discussed under the systematic uncertainty study in Sec.~\ref{sec:sys}~\ref{subsec:off_time_events}.
The \mbox{ToF-F} detector is not used further in the track~\mbox{selection}. It could not be employed in particle identification as the \mbox{ToF-F} resolution is insufficient for particle separation in the energy range of interest. However, the specific energy loss, \dedx, measurements alone provide good discrimination between protons and pions, as shown in Fig.~\ref{fig:dedx}. A linear graphical cut is used to separate the two particle species.\\
The most restrictive requirement in the selection is that track momentum must be larger than 29.73~\GeVc (the value of this cut is explained below). This cut value takes into account the continuous energy loss due to ionization in the target and the recoil energy in quasi-elastic interactions. The ionization losses inside the target follow a Landau distribution. Simulations were made to examine this distribution for~30.92~\GeVc protons going through a 90-cm-long graphite target with a density of 1.83~g/$\mathrm{cm}^{3}$. Its median was found to be $E_\mathrm{ion} = 315$~\MeV.\\
Assuming elastic and quasi-elastic interactions with either the carbon nuclei or with the nucleons inside the nuclei, the kinetic energy of the recoiling particle is given by
\begin{eqnarray}
  T_{R} = \frac{|t|}{2M} = \frac{p^{2}\theta^{2}}{2M},
  \label{eq:tr}
\end{eqnarray}
where $t$ is the four-momentum transfer, $M$ is the rest mass of the recoiling particle, $p$ is the momentum of the beam, and $\theta$ is the scattering angle between the incoming and outgoing particle. In elastic and quasi-elastic interactions, the scattered particle stays intact with most of its energy, and an approximation of the four-momentum transfer is $t\sim p^{2}\theta^{2}$. From Eq.~\ref{eq:tr}, the kinetic energy of the recoiling particle is largest when the recoil energy is transferred to a single nucleon, that is a knock-out reaction. A value of~30~mrad, at the tail of the scattering angle distribution for tracks with energy above 30~\GeV, is assigned to~$\theta$. Thus, the calculated maximal kinetic energy of the recoiling particle in quasi-elastic events is~459~\MeV. The length of the T2K replica target amounts to two nuclear interaction lengths and one can expect, on average, two nuclear scatterings for a particle passing through the whole target length. Taking this into account, the expected energy range of an outgoing elastically or quasi-elastically scattered beam particle,~$E_\mathrm{out}$,~is 
\begin{eqnarray}
    E_\mathrm{out} \geq E_\mathrm{beam} - E_\mathrm{ion} - 2T_{R},
    \label{eq:Ecut}
\end{eqnarray}
where $E_\mathrm{beam}$ is the total energy of the beam particles. In terms of proton momenta, Eq.~\ref{eq:Ecut} transforms to~\textit{$p\geq29.73$}~\GeVc.\\
In elastic and quasi-elastic events, alongside the high-energy selection candidate, low-energy nuclear fragments, electrons, protons, neutrons and de-excitation photons can be produced.
 Those that exit the target and get reconstructed in the TPCs\footnote{Neutral particles cannot be detected with a TPC} are bent away from the beam line and the spectrometer itself by the strong magnetic field and do not reach the ToF detectors. Particle identification for such tracks is not feasible since their energy loss measurements lie in a crossover region between electrons and $\pi^{-}$ or protons and $\pi^{+}$. This is depicted in the low-momenta range in Fig.~\ref{fig:dedx_others} which shows the energy loss vs momentum distribution for tracks that are reconstructed together with a high-energy selection candidate. Energy loss of the selected candidate is not plotted, thus at larger momenta Fig.~\ref{fig:dedx_others} shows off-time beam particles, primarily $\pi^{+}$ and protons, that have not undergone production interaction inside the target and have passed through the TPCs. An example of such a track is shown in Fig.~\ref{fig:off_time}. In order to collect the maximum number of elastically and quasi-elastically scattered particles without damaging the purity of the selection, it is required that every reconstructed track in the same event as the selection candidate, if any, must have a momentum larger than the calculated 29.73~\GeVc cut value. Therefore, interactions are discarded in cases where low-energy particles are detected by the TPCs, even if there is also a high-energy proton track in the same event.
The numbers of remaining events and tracks in the course of the selection process are given in Tab.~\ref{tab:N}.
\begin{table}[ht]
\centering
\begin{multicols}{2}
    \begin{tabular}{cc}
    \hline \hline
     Event selection & ($10^{6}$) \\
     \hline
     Total & 1.235 \\
     T3 trigger & 0.965 \\
     BPD measurement & 0.790 \\
     $\mathrm{R}_\mathrm{upstream}$ & 0.766\\
     $\mathrm{R}_\mathrm{downstream}$ & 0.766\\
     \hline\hline
    \end{tabular}   
    \\
\columnbreak
    \begin{tabular}{cc}
    \hline \hline
    Track selection & ($10^{6}$) \\
     \hline
     Distance from target& 2.839\\
     Charge & 1.848 \\
     \mbox{ToF-F} hit & 0.313 \\
     Target exit point & 0.195 \\
     \dedx measurement & 0.176 \\
     \textit{$p\geq29.73$}~\GeVc & 0.108\\
     \hline\hline
    \end{tabular}
\end{multicols}
\caption{Number of events and tracks after consecutive application of the selection criteria. The notations $R_\textrm{upstream}$ and $R_\textrm{downstream}$ indicate the extrapolated beam track passed through the upstream and would have passed through the downstream target face in the case there were no interactions inside the target.}
\label{tab:N}
\end{table}

\begin{figure}
    \centering
    \includegraphics[scale=0.36]{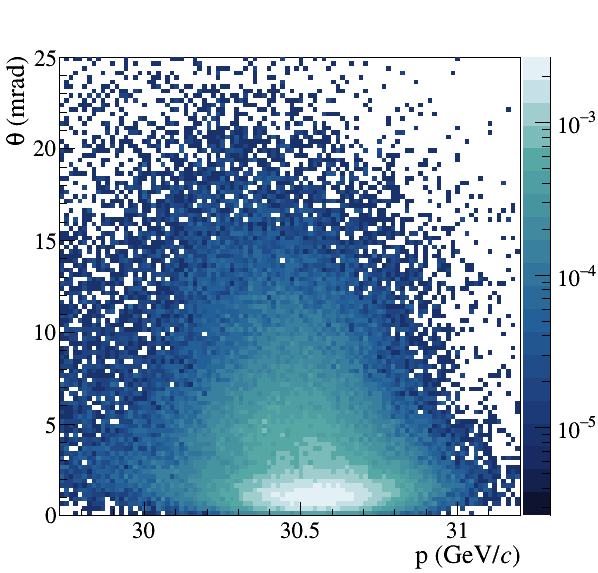}
\caption{Distribution of selected high-energy tracks in $\{ p, \theta \}$. The abscissa range starts at the 29.73~\GeVc momentum cut value. }
    \label{fig:phase_space}
\end{figure}

\begin{figure}
    \centering
    \includegraphics[scale=0.36]{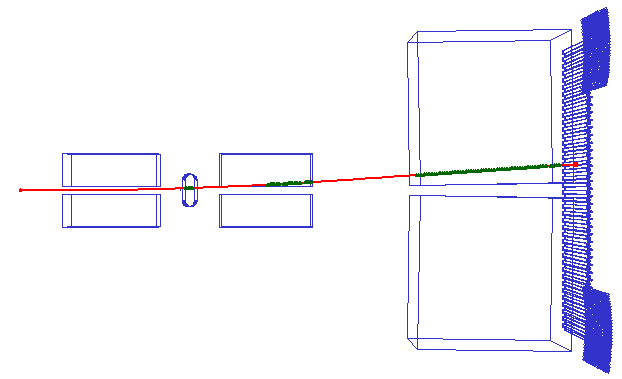}
\caption{Top view of a selected event - a single track going through \mbox{GTPC, VTPC-2} and \mbox{MTPC-L}. The green markers, 125 in total, represent the measured points along the track. The red line is the reconstructed particle trajectory starting from the target on the left to the \mbox{ToF-F} wall on the right. The track's reconstructed momentum is \textit{$p~=~30.34$}~\GeVc.}
    \label{fig:HEtrack}
\end{figure}

Phase space in $\{ p, \theta \}$ of selected high-energy tracks is plotted in Fig.~\ref{fig:phase_space}. The majority of tracks have polar angles below a few milliradians and momenta around 30.6~\GeVc, which corresponds to elastic scattering. The fractional momentum resolution for high-energy tracks having momenta larger than 29.73~\GeVc is $5\times 10^{-3}$. Since the precision of track momentum reconstruction depends on the number of points on track, it is worth mentioning that no restrictions on this number are applied. In this analysis, trajectories of selected particles can be divided into two topologies based on the segments they have in different TPCs. The majority of tracks leave clusters in \mbox{GTPC}, \mbox{VTPC-2} and \mbox{MTPC-L}. Less than 10$\%$ of the selected tracks only go through the GTPC and MTPC-L. For both topologies, track selection indirectly implies that the number of clusters in GTPC is the maximum possible, i.e 7, and in the \mbox{MTPC-L} their number is more than 20, out of 90 possible. Tracks that pass through the \mbox{VTPC-2} turn out to have more than 20 clusters in that chamber, out of a maximum of 72. A typical selected event, i.e., a single track passing through the \NASixtyOne spectrometer, is shown in Fig.~\ref{fig:HEtrack}. This track has a total of 125 measured points, hits the \mbox{ToF-F} wall, and has momentum of 30.34~\GeVc.
\subsection{Survival probability}
The probability, $P_\mathrm{surv}$, that a beam particle avoids a production interaction inside the 90-cm-long target and the production cross section $\sigma_\mathrm{prod}$ are related via:
\begin{eqnarray}
    P_\mathrm{surv} = \frac{\text{Number of selected tracks}}{\text{Number of beam particles}} = e^{-L \cdot n \cdot \sigma_\mathrm{prod}},
    \label{eq:Psurv}
\end{eqnarray}
where $L$ is the target length and $n$ is the number density of the target nuclei. Following the event and track selection procedures described in Secs.~\ref{sec:ana}~\ref{subsec:event_sel}~and~\ref{sec:ana}~\ref{subsec:track_sel} the number of selected beam particles is~766,164 and the number of selected tracks is 108,378. Before extracting the production cross section, the probability $P_\mathrm{surv}$ is corrected for various effects. 
\subsection{Correction factors}
\label{subsec:MC_corr}
Two separate corrections are applied to the probability $P_\mathrm{surv}$: a Monte Carlo-based acceptance correction and a data-based ToF-F efficiency correction. Their product gives the total correction factor.\\
Approximately 6~$\times 10^{6}$ events were simulated using \GeantFour version 10.4.p03~\cite{Agostinelli:2002hh, Allison:2006ve, Allison:2016ve}. The beam profile from data was used to generate the Monte Carlo beam profile. The Monte Carlo correction is defined~as
\begin{eqnarray}
   C^{MC} = \frac{\text{Number of non-production simulated events}}{\text{Number of selected reconstructed events}},
   \label{eq:MCcorr}
\end{eqnarray}
where the numerator is the number of simulated beam protons that pass thought the whole target length and do not produce new hadrons inside it. The denominator is the number of reconstructed Monte Carlo events that are accepted as elastic or quasi-elastic by the same selection procedure that is applied to the data. Hence, by construction, any variation in the event and track selection has an impact on the magnitude of the MC correction factor. On the other hand, in the simulation one must accurately reproduce the data-taking environment, including the detector geometry. Thereby, the reconstruction algorithm in the Monte Carlo and the data case is run under similar conditions and a proper estimate of the MC correction can be obtained. Furthermore, the MC correction is sensitive to changes related to the generation of particle interactions inside the target, which is model dependent. All of the above-mentioned biases are addressed in Secs.~\ref{sec:sys} and \ref{sec:model}. The calculated MC correction for the reference Monte Carlo sample (QBBC physics list from \GeantFour) is $C^{MC}=1.035$.\\
The \mbox{ToF-F} efficiency correction factor relies on the calculation of the efficiency of the \mbox{ToF-F} scintillator bars from the data. In the Monte Carlo, the response of the ToF detectors is not simulated. Simply if a simulated track passes through the \mbox{ToF-F} wall, a \mbox{ToF-F} hit is assigned to it. The procedure to calculate the \mbox{ToF-F} efficiency is described in~\cite{Abgrall:2019p}. The efficiency of each scintillator bar in the \mbox{ToF-F} wall is defined as the ratio of the number of tracks that hit that bar 
to the number of tracks that reach the end of MTPCs and are extrapolated to the \mbox{ToF-F} $z$-position of that scintillator bar. Then, the \mbox{ToF-F} efficiency in the phase space of the selected tracks is calculated as
\begin{eqnarray}
   \epsilon^\mathrm{tof} = \frac{n_\mathrm{sel}}{\sum_{s} n_\mathrm{sel}^{s}/\epsilon^{s}},
\label{eq:ToFeff}
\end{eqnarray}
where $n_\mathrm{sel}$ is the total number of selected data tracks, $n_\mathrm{sel}^{s}$ is the number of selected data tracks that hit bar \textit{s} and $\epsilon^{s}$ is the efficiency of this bar \textit{s}. The summation goes over every active bar. The sample of all selected tracks produced signals in around 10 scintillator bars. The estimated \mbox{ToF-F} efficiency is around~96$\%$\footnote{The efficiency value is reduced by a quality cut on the timing difference between the signals of the top and bottom PMTs of each scintillator bar. The cut is implemented in order to improve separation between particles for the mass squared particle~identification. Even though this particle identification method is not applicable in the current analysis, the quality cut remains in the assessment of the \mbox{ToF-F} efficiency.}. Its uncertainty serves as an estimate of the \mbox{ToF-F} systematic uncertainty.

\subsection{Production cross section}
Using Eq.~\ref{eq:Psurv} for the survival probability and applying the correction factors from Eqs.~\ref{eq:MCcorr} and~\ref{eq:ToFeff}, the production cross section $\sigma_\mathrm{prod}$ becomes
\begin{eqnarray}
     \sigma_\mathrm{prod} = -ln \left ( P_\mathrm{surv} \times C^{MC} \times \frac{1}{\epsilon^\mathrm{tof}} \right )/(L\cdot N_{A}\cdot\rho/\mu),
    \label{eq:xs}
\end{eqnarray}
where the number density, $n=N_{A} \cdot \rho/\mu$, is given in terms of Avogadro's number $N_{A}$, the target material density and molar mass, $\rho$ and $\mu$. 
\section{Systematic uncertainties}
\label{sec:sys}
As a result of many cross-checks and a broad overview of previous analyses within \NASixtyOne, several sources of systematic uncertainty were identified. In the current section, experimental systematic uncertainties are discussed, while the physics model uncertainty is presented in Sec.~\ref{sec:model}. The magnitude of each systematic effect is calculated as the deviation from unity of the ratio of the recalculated production cross section to the standard, nominal one.

\subsection{Target density}
The target density enters into the production cross-section calculation via the number density as shown in~Eq.~\ref{eq:xs}. The reported target density is $1.83 \pm 0.01$ g/$\mathrm{cm}^{3}$. The uncertainty of the target density is propagated to the production cross-section result. The corresponding effect is $\pm 0.6\%$.

\subsection{Backward track extrapolation}
The target position and tilt, discussed in Sec.~\ref{sec:ana}~\ref{subsec:Tpos}, have a finite precision that can induce biases when extrapolating TPC tracks backwards to the target surface. To study this bias, in the analysis codes the target position and tilt were changed within the calibration uncertainties. The data were reprocessed resulting in supplemental estimates of the survival probability and thus production cross-section values. The effects of the target position and tilt change in each axis and plane are added in quadrature to form the total uncertainty, which was found to be around $ \pm 0.1\%$.

\subsection{Beam spot size on upstream target face}
Since many re-interactions can take place inside the 90-cm-long replica target, tracks that hit the outer edge of the target are more likely to be scattered outside of the target before reaching its downstream end than those that hit the target center. To study this effect, for both data and Monte Carlo, the radius of the beam profile on the target front face is reduced from 1.28~cm to 0.65~cm, cutting down 10$\%$ of the selected events. Both the survival probability $P_\textrm{surv}$ and the MC correction $C^{MC}$ are recalculated and lead to additional production cross-section estimates. The assigned systematic uncertainty is~$-0.2\%$.

\subsection{Particle identification of low-energy products}
In elastic and quasi-elastic events, besides the high-energy proton, low-energy nuclear fragments, electrons, protons, neutrons and de-excitation photons can be produced. In reality, the species of the above-mentioned charged particles cannot be resolved (see Fig.~\ref{fig:dedx_others}) and therefore the interactions in which they are produced are not selected. However, simulations provide particle identification (PID) that can be consulted when calculating the Monte Carlo correction factor described in Sec.~\ref{sec:ana}~\ref{subsec:MC_corr}. To facilitate this study, matching of simulated and reconstructed tracks is performed and one knows the true particle type corresponding to each reconstructed track in the MC sample. Then, reconstructed MC events, in which a high-energy proton is produced alongside low-energy particles, such as protons, electrons, and nuclear fragments, can be flagged as elastic or quasi-elastic. Given the high-energy proton in such flagged events meets all other selection criteria, those events, as well as selected events with a single reconstructed high-energy proton track, will be used in the estimation of the MC correction factor. Their combined number will become the denominator in Eq.~\ref{eq:MCcorr}. Overall, the described procedure mimics the case, where PID is possible for all charged particles produced in an event. Using Monte Carlo PID, the MC correction factor is recalculated and so is the production cross section. The systematic uncertainty attributed to particle identification of low-energy products is $ - 0.4\%$.

\subsection{Proton loss}
Beam protons that have scattered elastically or quasi-elastically in the target may miss the \mbox{ToF-F} wall or re-interact in the detector. Removal of the requirement for a ToF-F hit adds these tracks to the selected sample. To account for off-time beam particles surviving the interactions in the target, for any event in which more than one high-energy track is reconstructed, a single track is accepted. Additionally, to guarantee high-quality track reconstruction and momentum fit, selected tracks in this study must have a segment in VTPC-2. With no requirement for a \mbox{ToF-F} detection, but with a \mbox{VTPC-2} measurement, the survival probability $P_\textrm{surv}$ and the MC correction $C^{MC}$ are recomputed. The corresponding production cross-section value is compared to the standard result. A $ - 0.1\%$ effect is assigned to hadron loss.

\subsection{Off-time events}
\label{subsec:off_time_events}
When the triggered beam particle undergoes an elastic or quasi-elastic interaction, but there is also an off-time event that produces new hadrons in the target (an off-time production event), the surviving beam particle will not be selected. The reason is that the products of the two interaction types cannot be separated. However, the probability of such a combination of events can be estimated based on the number of cases where both the triggered and the off-time beam particles scatter elastically or quasi-elastically. This is possible since the probabilities of any type of interaction in the target is independent of whether the beam particle is counted by the trigger system or not. The survival probability estimate from this analysis is used. The number of beam particles avoiding production interactions in the replica target but accompanied by products of an off-time beam particle production interaction is estimated and added to the total number of selected tracks. This alters the survival probability in Eq.~\ref{eq:xs}. The production cross section is recalculated, compared to the nominal result and the systematic effect is found to be $ - 0.8\%$. 

\subsection{ToF efficiency uncertainty}
Particles may also lack a recorded \mbox{ToF-F} hit due to the inefficiency of the \mbox{ToF-F} detector. Also, \mbox{ToF-F} efficiency is used as a correction factor (see Sec.~\ref{sec:ana}~\ref{subsec:MC_corr}), and so influences the production cross-section result. For these reasons, the uncertainty of the calculated \mbox{ToF-F} efficiency is propagated to the production cross-section uncertainty. The corresponding systematic effect is $ \pm 0.3\%$.

\subsection{Reconstruction}
Potential differences between the spectrometer description in the simulation and the real detector geometry would affect track reconstruction and could bias the MC correction factor. To address such an effect, the \mbox{VTPC-2} position used for the reconstruction of simulated tracks is shifted by $\pm$ 0.2 mm in~$x$ and $\pm$ 0.3 mm in the $y$-direction. Hence, this uncertainty source reflects the impact that small shifts in the detector positions have on the reconstruction of tracks in the MC chain. The size of the shifts are chosen to be much larger than any observed alignment and residual effects in the calibration of the data. The other detectors are not moved since the selected high-energy tracks leave no clusters in VTPC-1, the GTPC covers the forward region, and the MTPCs are well outside of the magnetic field and are not used for momentum determination. The Monte Carlo correction factor and the production cross section are recalculated for each of the produced four MC samples. Then their deviations from the nominal production cross-section value are added in quadrature, taking into account the corresponding sign, to form the asymmetric systematic uncertainty due to reconstruction effects. The resulting fractional uncertainty is~$ _{-0.8}^{+0.5}\%$.


\subsection{Track momentum cut}
As discussed in Sec.~\ref{sec:ana}~\ref{subsec:track_sel}, the most constraining selection criterion is the momentum requirement \mbox{$p \geq 29.73$ \GeVc}. On one side, the track momentum reconstruction depends on the precision of TPC track reconstruction and backward extrapolation. These two factors are separately studied. On the other hand, the cut value itself depends on the beam momentum resolution. It is reported that the beam momentum spread is less than 1$\%$~\cite{Abgrall:2014fa}. This spread is in turn propagated to the cut value uncertainty. Then the cut value is varied within its estimated uncertainty and the track selection is repeated. The change in track selection reflects on both the survival probability $P_\textrm{surv}$ and the MC correction $C^{MC}$. The corresponding systematic effect is $ \pm 0.2\%$. In this evaluation, no uncertainty is attributed to the value of the energy loss due to ionization inside the target.
%
\section{Physics model uncertainty}
\label{sec:model}
\begin{figure}[t!]
    \begin{subfigure}[t]{0.99\textwidth}
        \centering
        \includegraphics[width=\linewidth]{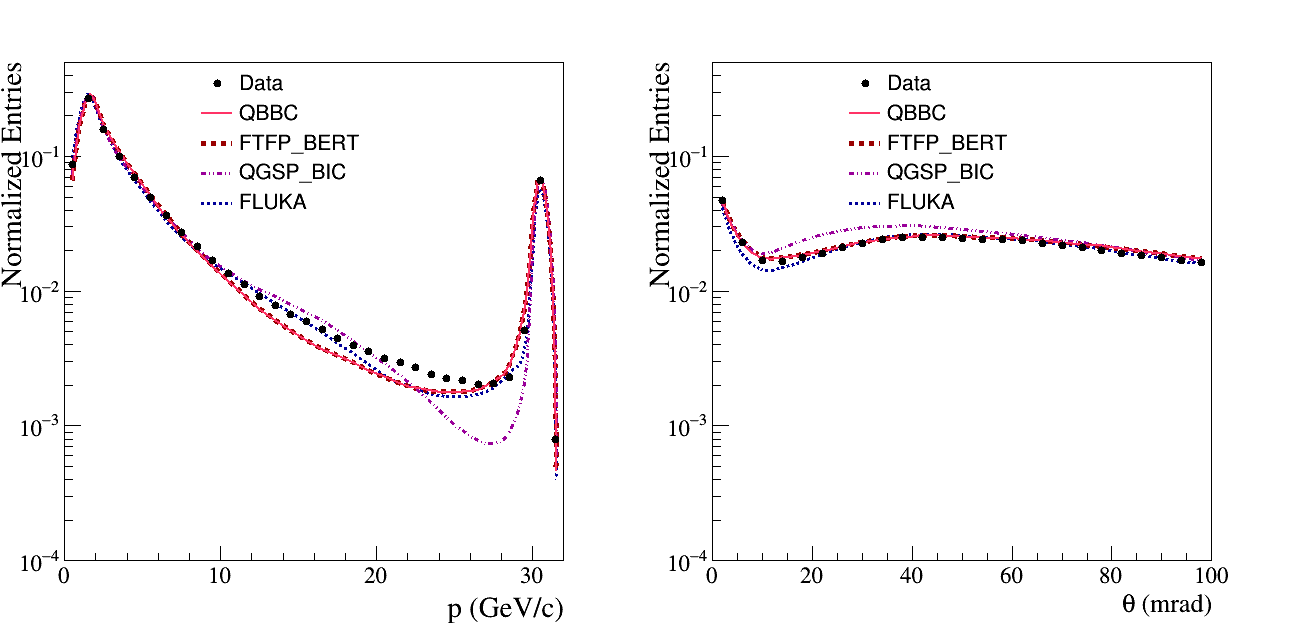}\\
        \caption{}
        \label{fig:allTrks}
    \end{subfigure}
    \begin{subfigure}[t]{0.99\textwidth}
        \centering
        \includegraphics[width=\linewidth]{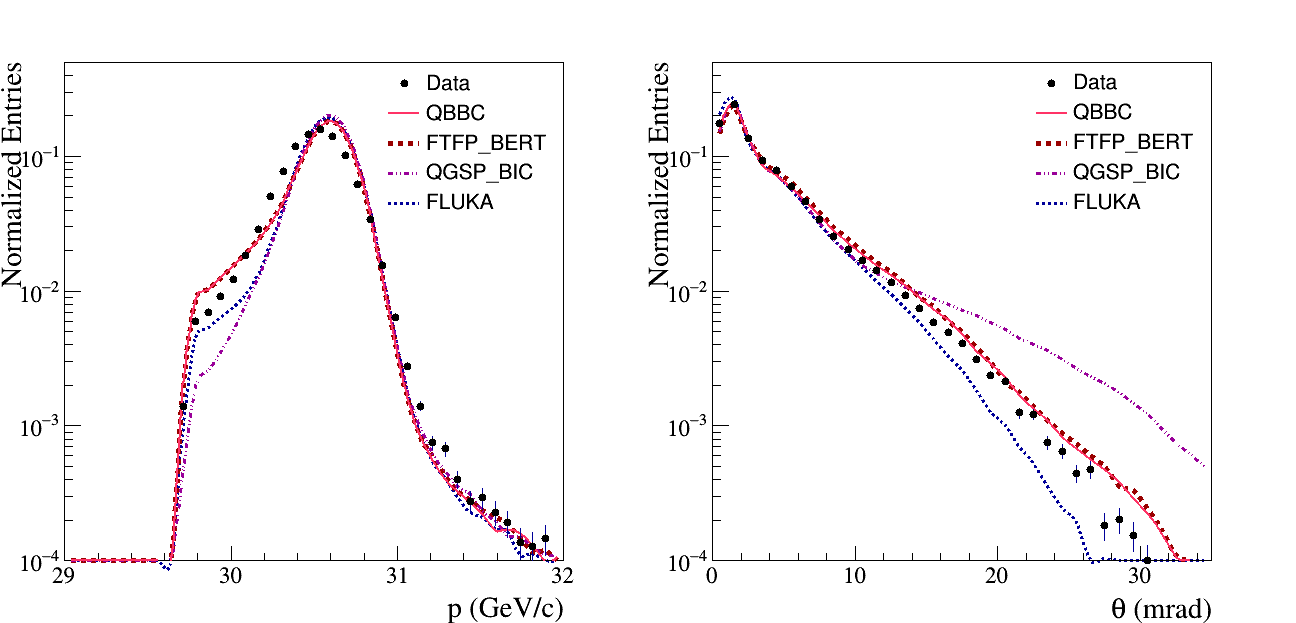}\\
        \caption{}
        \label{fig:heTrks}
    \end{subfigure}
\caption{Normalized momentum and polar angle distributions for all reconstructed tracks having more than 30 points in the detector (\subref{fig:allTrks}), and for selected high-energy tracks (\subref{fig:heTrks}). Presented are distributions for the three \GeantFour physics lists (QBBC, FTFP\textunderscore BERT, QGSP\textunderscore BIC) and \mbox{\Fluka 2011.2c.5.} The color scheme is given on each plot. The sharp cut in the momentum distribution on the left-hand plot (\subref{fig:heTrks}) corresponds to the \textit{$p\geq29.73$} \GeVc track selection requirement.\\\\\\\\}
\label{fig:comparison}
\end{figure}

Variations in Monte Carlo modeling of physics processes can lead to different MC correction factors. To estimate the corresponding uncertainty, three \GeantFour physics lists were used: QBBC, FTFP\textunderscore BERT and QGSP\textunderscore BIC. The treatment of elastic processes for FTFP\textunderscore BERT and QGSP\textunderscore BIC is identical, but it differs for QBBC. In general, modeling of inelastic, and so of quasi-elastic, interactions changes within each physics list. However, in the range of interest, which is above \textit{$p\geq29.73$}~\GeVc, QBBC and FTFP\textunderscore BERT both use the Fritiof parton model. The reference Monte Carlo model used in this analysis is the one employed in QBBC. The other two physics lists are used to evaluate the physics model systematic uncertainty. A separate Monte Carlo set using the \mbox{\Fluka 2011.2c.5} package~\cite{Ferrari:2005zk, Bohlen:2014buj, Battistoni:2007zzb} was also prepared. Normalized momentum, $p$, and polar angle, $\theta$, distributions for reconstructed data and MC events are shown in Fig.~\ref{fig:comparison}. For Fig.~\ref{fig:allTrks}, all tracks that have more than 30 points in the TPCs are processed, while Fig.~\ref{fig:heTrks} is for tracks that pass all of the selection criteria. Table \ref{tab:MCxs} presents production cross sections calculated using the survival probability extracted from the data, applying Monte Carlo corrections calculated with each of the three \GeantFour lists as well as with \mbox{\Fluka 2011.2c.5.} Previous \NASixtyOne production cross-section analyses used a \GeantFour-based MC correction, while the latest \NASixtyOne results for hadron production in the T2K replica target rely on \mbox{\Fluka 2011.2c.5} simulations for the MC correction of particle yields. Furthermore, the T2K neutrino flux simulations feature \Fluka as a generator of the hadronic interactions inside the target. For these reasons, and as a cross-check, the additional \mbox{\Fluka 2011.2c.5} Monte Carlo was produced and corresponding MC correction was applied to the data. The \Fluka-corrected production cross section is in best agreement with the FTFP\textunderscore BERT one. Overall, production cross-section estimates with MC corrections from the three \GeantFour models and the \FlukaEleven model show good agreement. The maximum magnitude of the uncertainty due to modeling of physics interactions is found for the~QGSP\textunderscore BIC list, and it is $ - 0.4\%$.
\begin{table}[ht]
\centering
\begin{tabular}{cccc}
\hline \hline
     Physics model  & MC correction &  Measured  & 1 -  $\sigma_\mathrm{prod}^\mathrm{ref}$/$\sigma_\mathrm{prod}$ \\
     used for correction &$C^{MC}$ & $\sigma_\mathrm{prod}$ (mb) & ($\%$)\\
    \hline
     QBBC & 1.035 & 227.6 & - \\
     FTFP\textunderscore BERT & 1.036 & 227.5 & - 0.04\\
     QGSP\textunderscore BIC & 1.042 & 226.8 & - 0.4\\
     \Fluka 2011.2c.5  &1.037 & 227.4 & - 0.09\\
\hline \hline
\end{tabular}
\caption{Monte Carlo correction, the measured production cross section obtained using the corresponding MC correction and assigned fractional uncertainty using \GeantFour version 10.4.p03 and \mbox{\Fluka 2011.2c.5} interaction generators. QBBC, FTFP\textunderscore BERT, QGSP\textunderscore BIC are the employed three physics lists from \GeantFour. The reference MC model is QBBC.}
\label{tab:MCxs}
\end{table}
\section{Results and comparisons}
\label{sec:res}
\begin{figure}
    \centering
    \includegraphics[scale=0.35]{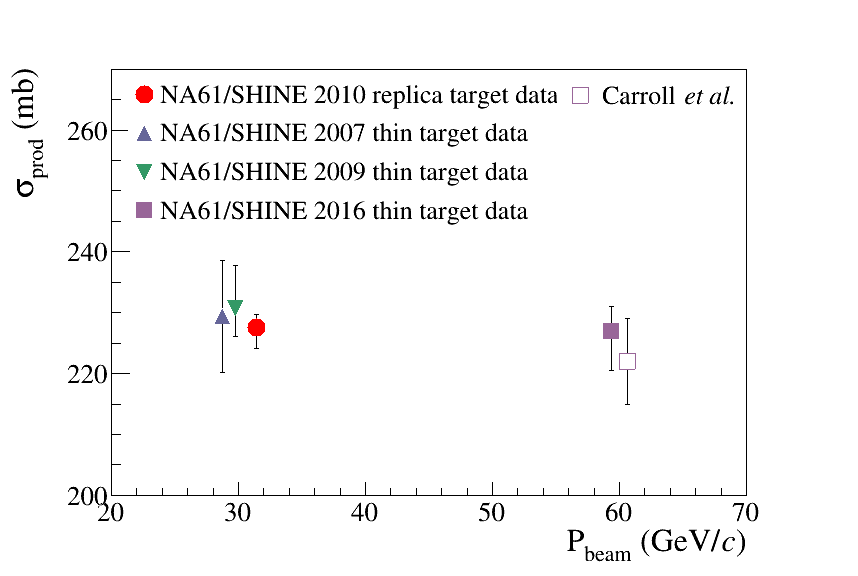}
\caption{The result of this analysis compared to the production cross-section measurements for \mbox{p + C} interactions at different beam momenta. Alongside the \NASixtyOne thin-target results, production cross section by \mbox{Carroll \textit{et al.}} at 60~\GeVc~\cite{Carroll:1978hc} is shown.}
    \label{fig:xs_comp}
\end{figure}

A production cross section for 30.92~\GeVc protons on carbon was estimated based on the number of beam particles that pass through the 90-cm-long T2K replica target, interact elastically or quasi-elastically inside it, and leave a track in the spectrometer. The result, including statistical, systematic, and model uncertainties, is
\begin{eqnarray}
    \mathrm{\sigma_{prod} = 227.6 \pm 0.8 (stat)  \: _{-3.2}^{+1.9} (sys) \: {-0.8} (mod) \: mb} .  
\end{eqnarray}
A list of all identified uncertainty sources and their fractional magnitudes is given in Tab.~\ref{tab:uncert}. The reported production cross-section measurement is in agreement with previous \NASixtyOne results for \mbox{p + C} at 31~\GeVc~\cite{Abgrall:2011ae,Abgrall:2015hmv}, while providing better precision, as shown in Fig.~\ref{fig:xs_comp} and  Tab.~\ref{tab:xs}. For comparison, the production cross-section measurements for \mbox{p + C} at 60~\GeVc by \NASixtyOne \footnote{ Part of \NASixtyOne's hadron production measurements for the Fermilab neutrino experiments}~\cite{Aduszkiewicz:2019xna} and by \mbox{Carroll \textit{et al.}}~\cite{Carroll:1978hc} are also given.
\begin{table}
\centering
\begin{tabular}{lcc}
\hline \hline
     Uncertainty source & Fractional size ($\%$) \\
    \hline
    Target density & $\pm$0.6\\
    Backward track extrapolation & $\pm$0.1\\
    Beam spot size on upstream target face & -0.2\\
    PID of low-energy products & -0.4\\
    Proton loss & -0.1\\
    Off-time events & -0.8\\
    ToF efficiency & $\pm$0.3\\
    Reconstruction & $_{-0.8}^{+0.5}$\\
    Track momentum cut & $\pm$0.2\\
    Physics model & -0.4\\
    Statistical &  $\pm$0.4 \\
    \hline \hline
\end{tabular}
\caption{\label{tab:uncert} Uncertainty sources considered in the current production cross-section measurement and their fractional size. The discussion of each item is given in Secs.~\ref{sec:sys} and \ref{sec:model}.}
\end{table}
\begin{table}
\centering
\begin{tabular}{cccc}
\hline \hline
     Experiment & $p_\mathrm{beam}$ (\GeVc) & $\sigma_\mathrm{prod}$ (mb) & $\Delta_\mathrm{total}$ (mb) \\
    \hline
     NA61/SHINE 2010 & 31 & 227.6  & $_{-3.4}^{+2.1}$\\
     NA61/SHINE 2007 & 31 & 229.3 & $\pm$ 9.2  \\
     NA61/SHINE 2009 & 31 & 230.7 & $_{-4.6}^{+7.0}$\\
     NA61/SHINE 2016 & 60 & 226.9 & $_{-6.4}^{+4.1}$\\
     Carroll \textit{et al.} & 60 & 222 & $\pm$ 7  \\
\hline \hline
\end{tabular}
\caption{Result of this analysis compared to the production cross-section thin-target measurements by \NASixtyOne~\cite{Abgrall:2011ae,Abgrall:2015hmv,Aduszkiewicz:2019xna} and \mbox{Carroll \textit{et al.}}~\cite{Carroll:1978hc} for \mbox{p + C} interactions at different beam momenta. The total uncertainty~($\Delta_\mathrm{total}$) is the statistical,
systematic and model uncertainties added in quadrature.}
\label{tab:xs}
\end{table}
\section{Summary}
\label{sec:summary}
This paper presents a direct method for a production cross-section measurement using the attenuation of beam particles in a thick medium. The production cross section of 30.92~\GeVc protons on carbon has been obtained using data collected by the \NASixtyOne collaboration with the 90-cm-long T2K replica target. About 2$\%$ total uncertainty is estimated by careful studies of statistical, systematic and model uncertainties. The resulting production cross section is in agreement with previous measurements and has the smallest total uncertainty. It will allow for a more accurate hadron interaction rate re-weighting of the T2K neutrino flux prediction, as its estimation is based on the simultaneous selection of both elastic and quasi-elastic processes inside the T2K replica target.
\section*{Acknowledgments}
We would like to thank the CERN EP, BE, EN and HSE Departments for the
strong support of \NASixtyOne.

This work was supported by
the Hungarian Scientific Research Fund (grant NKFIH 123842\slash123959),
the Polish Ministry of Science
and Higher Education (grants 667\slash N-CERN\slash2010\slash0,
NN\,202\,48\,4339, NN\,202\,23\,1837 and DIR\slash WK\slash 2016\slash 2017\slash 10-1), the National Science Centre Poland (grants~2014\slash14\slash E\slash ST2\slash00018, 2014\slash15\slash B\slash ST2 \slash\- 02537 and
2015\slash18\slash M\slash ST2\slash00125, 2015\slash 19\slash N\slash ST2 \slash01689, 2016\slash23\slash B\slash ST2\slash00692,
2017\slash\- 25\slash N\slash\- ST2\slash\- 02575,
2018\slash 30\slash A\slash ST2\slash 00226,
2018\slash 31\slash G\slash ST2\slash 03910),
the Russian Science Foundation, grant 16-12-10176 and 17-72-20045,
the Russian Academy of Science and the
Russian Foundation for Basic Research (grants 08-02-00018, 09-02-00664
and 12-02-91503-CERN),
the Russian Foundation for Basic Research (RFBR) funding within the research project no. 18-02-40086,
the National Research Nuclear University MEPhI in the framework of the Russian Academic Excellence Project (contract No.\ 02.a03.21.0005, 27.08.2013),
the Ministry of Science and Higher Education of the Russian Federation, Project "Fundamental properties of elementary particles and cosmology" No 0723-2020-0041,
the European Union's Horizon 2020 research and innovation programme under grant agreement No. 871072,
the Ministry of Education, Culture, Sports,
Science and Tech\-no\-lo\-gy, Japan, Grant-in-Aid for Sci\-en\-ti\-fic
Research (grants 18071005, 19034011, 19740162, 20740160 and 20039012),
the German Research Foundation (grant GA\,1480/8-1), the
Bulgarian Nuclear Regulatory Agency and the Joint Institute for
Nuclear Research, Dubna (bilateral contract No. 4799-1-18\slash 20),
Bulgarian National Science Fund (grant DN08/11), Ministry of Education
and Science of the Republic of Serbia (grant OI171002), Swiss
Nationalfonds Foundation (grant 200020\-117913/1), ETH Research Grant
TH-01\,07-3 and the Fermi National Accelerator Laboratory (Fermilab), a U.S. Department of Energy, Office of Science, HEP User Facility managed by Fermi Research Alliance, LLC (FRA), acting under Contract No. DE-AC02-07CH11359 and the IN2P3-CNRS (France).

\bibliography{na61References}
\newpage
{\Large The \NASixtyOne Collaboration}
\bigskip
\begin{sloppypar}

\noindent
A.~Acharya$^{\,9}$,
H.~Adhikary$^{\,9}$,
A.~Aduszkiewicz$^{\,15}$,
K.K.~Allison$^{\,25}$,
E.V.~Andronov$^{\,21}$,
T.~Anti\'ci\'c$^{\,3}$,
V.~Babkin$^{\,19}$,
M.~Baszczyk$^{\,13}$,
S.~Bhosale$^{\,10}$,
A.~Blondel$^{\,4}$
% ...
\end{sloppypar}

\noindent
% ...
$^{3}$~Ru{\dj}er Bo\v{s}kovi\'c Institute, Zagreb, Croatia\\
$^{4}$~LPNHE, University of Paris VI and VII, Paris, France\\
% ...
$^{9}$~Jan Kochanowski University in Kielce, Poland\\
$^{10}$~Institute of Nuclear Physics, Polish Academy of Sciences, Cracow, Poland\\
% ...
$^{13}$~AGH - University of Science and Technology, Cracow, Poland\\
% ...
$^{15}$~University of Warsaw, Warsaw, Poland\\
% ...
$^{19}$~Joint Institute for Nuclear Research, Dubna, Russia\\
% ...
$^{21}$~St. Petersburg State University, St. Petersburg, Russia\\
% ...
$^{25}$~University of Colorado, Boulder, USA\\
% ...

\end{document}